\documentclass[12pt, preprint]{aastex}
\usepackage{geometry} 
\usepackage{gensymb}
\usepackage[]{natbib}

\slugcomment{Accepted to ApJ}

\shorttitle{MOST observations of $\delta$ Ori} 
\shortauthors{Pablo et al.} 

\begin{document}

\title{A Coordinated X-ray and Optical Campaign of the Nearest Massive Eclipsing Binary,
$\delta$ Orionis Aa: III. Analysis of Optical  Photometric (\textit{MOST}) and Spectroscopic (Ground Based) Variations}
\author{Herbert Pablo\altaffilmark{1},
Noel D. Richardson\altaffilmark{1}, 
Anthony F. J. Moffat\altaffilmark{1},
Michael Corcoran\altaffilmark{2,3},
Tomer Shenar\altaffilmark{4},
Omar Benvenuto\altaffilmark{5,6},
Jim Fuller\altaffilmark{7,8},
Ya\"el Naz\'e\altaffilmark{9},
Jennifer L. Hoffman\altaffilmark{10},
Anatoly Miroshnichenko\altaffilmark{11},
Jes{\'us} Ma\'iz Apell\'aniz\altaffilmark{12},
Nancy Evans\altaffilmark{13},
Thomas Eversberg\altaffilmark{14},
Ken Gayley\altaffilmark{15},
Ted Gull\altaffilmark{16},
Kenji Hamaguchi\altaffilmark{2},
Wolf-Rainer Hamann\altaffilmark{4},
Huib Henrichs\altaffilmark{17},
Tabetha Hole\altaffilmark{18},
Richard Ignace\altaffilmark{18},
Rosina Iping\altaffilmark{3},
Jennifer Lauer\altaffilmark{13},
Maurice Leutenegger\altaffilmark{8},
Jamie Lomax\altaffilmark{19},
Joy Nichols\altaffilmark{13},
Lida Oskinova\altaffilmark{4},
Stan Owocki\altaffilmark{20},
Andy Pollock\altaffilmark{21},
Christopher M. P. Russell\altaffilmark{22,23},
Wayne Waldron\altaffilmark{24},
Christian Buil\altaffilmark{25},
Thierry Garrel\altaffilmark{26},
Keith Graham\altaffilmark{27},
Bernard Heathcote\altaffilmark{28},
Thierry Lemoult\altaffilmark{29},
Dong Li\altaffilmark{30},
Benjamin Mauclaire\altaffilmark{31},
Mike Potter\altaffilmark{32},
Jose Ribeiro\altaffilmark{33},
Jaymie Matthews\altaffilmark{34},
Chris Cameron\altaffilmark{35},
David Guenther\altaffilmark{36},
Rainer Kuschnig\altaffilmark{34,37},
Jason Rowe\altaffilmark{38},
Slavek Rucinski\altaffilmark{39},
Dimitar Sasselov\altaffilmark{40}, and 
Werner Weiss\altaffilmark{37}
}

\altaffiltext{1}{D\'epartement de physique and Centre de Recherche en Astrophysique du Qu\'ebec (CRAQ), Universit\'e de Montr\'eal, C.P. 6128, Succ.~Centre-Ville, Montr\'eal, Qu\'ebec, H3C 3J7, Canada; hpablo@astro.umontreal.ca, richardson@astro.umontreal.ca}
\altaffiltext{2}{CRESST and X-ray Astrophysics Laboratory, NASA/GSFC, Greenbelt, MD 20771, USA}
\altaffiltext{3}{Universities Space Research Association, 7178 Columbia Gateway Drive, Columbia, MD 21046, USA}
\altaffiltext{4}{Institut f\"{u}r Physik und Astronomie, Universit\"{a}t Potsdam, Karl-Liebknecht-Str. 24/25, 14476, Potsdam, Germany}
\altaffiltext{5}{Facultad de Ciencias Astron\'{o}micas y Geof\'{i}sicas, Universidad Nacional de La Plata, 1900 La Plata, Buenos Aires, Argentina} 
\altaffiltext{6}{Instituto de Astrofs\'{i}ica de La Plata (IALP), CCT-CONICET-UNLP. Paseo del Bosque S/N (B1900FWA), La Plata, Argentina}
\altaffiltext{7}{TAPIR, Walter Burke Institute for Theoretical Physics, Mailcode 350-17, California Institute of Technology, Pasadena, CA 91125, USA}
\altaffiltext{8}{Kavli Institute for Theoretical Physics, Kohn Hall, University of California, Santa Barbara, CA 93106, USA}
\altaffiltext{9}{FNRS D\'epartement AGO, Universit\'e de Li\`ege, All\'ee du 6 Aout 17, Bat. B5C, 4000, Li\`ege, Belgium}
\altaffiltext{10}{Department of Physics \& Astronomy, University of Denver, 2112 East Wesley Avenue, Denver, CO 80208, USA}
\altaffiltext{11}{Department of Physics and Astronomy, University of North Carolina at Greensboro, Greensboro, NC 27402-6170, USA}
\altaffiltext{12}{Centro de Astrobiolog{\'\i}a (CSIC-INTA), ESAC Campus, P.O. Box 78,  28691 Villanueva de la Cañada, Madrid, Spain}

\altaffiltext{13}{Smithsonian Astrophysical Observatory, 60 Garden St., Cambridge, MA 02138, USA}
\altaffiltext{14}{Schn\"{o}rringen Telescope Science Institute, Waldbr\"{o}l, Germany}
\altaffiltext{15}{Department of Physics and Astronomy, University of Iowa, Iowa City, IA 52242}

\altaffiltext{16}{Astrophysics Science Division, NASA Goddard Space Flight Center, Greenbelt, MD 20771, USA}

\altaffiltext{17}{Astronomical Institute "Anton Pannekoek", University of Amsterdam, Science Park 904, 1098 XH Amsterdam, The Netherlands}
\altaffiltext{18}{Department of Physics \& Astronomy, East Tennessee State University, Box 70652, Johnson City, TN 37614, USA}

\altaffiltext{19}{HL Dodge Department of Physics and Astronomy, University of Oklahoma, Norman, OK, USA}
\altaffiltext{20}{Bartol Research Institute, University of Delaware, Newark, DE 19716, USA}
\altaffiltext{21}{European Space Agency, Apartado 78, Villanueva de la Canada, E-28691 Madrid, Spain}
\altaffiltext{22}{X-ray Astrophysics Lab, Code 662, NASA Goddard Space Flight Center, Greenbelt, MD
20771 USA}
\altaffiltext{23}{Oak Ridge Associated Universities (ORAU), Oak Ridge, TN 37831 USA}

\altaffiltext{24}{Eureka Scientific Inc., 2452 Dellmer Street, Suite 100, Oakland, CA 94602, USA}
\altaffiltext{25}{Castanet Tolosan Observatory, 6 place Cl\'emence Isaure, 31320, Castanet Tolosan, France}
\altaffiltext{26}{Observatoire de Juvignac, 19 avenue du Hameau du Golf, 34990 Juvignac, France}
\altaffiltext{27}{The ConVento Group}
\altaffiltext{28}{Barfold Observatory, Glenhope, Victoria 3444, Australia}
\altaffiltext{29}{Chelles Observatory, 23 avenue H{\'e}nin, 77500 Chelles, France}
\altaffiltext{30}{Jade Observatory, Jin Jiang Nan Li, He Bei District, 300251 Tianjin, China}
\altaffiltext{31}{Observatoire du Val de lÕArc, route de Peynier, 13530 Trets, France}
\altaffiltext{32}{3206 Overland Ave, Baltimore, MD 21214 USA}
\altaffiltext{33}{Observatorio do Instituto Geografico do Exercito, Lisboa, Portugal}
\altaffiltext{34}{Department of Physics and Astronomy, University of British Columbia, 6224 Agricultural Road, Vancouver, BC V6T 1Z1, Canada}
\altaffiltext{35}{Department of Mathematics, Physics \& Geology, Cape Breton University, 1250 Grand Lake Road, Sydney, Nova Scotia, Canada, B1P 6L2}
\altaffiltext{36}{Institute for Computational Astrophysics, Dept. of Astronomy and Physics, St Mary's University Halifax, NS B3H 3C3, Canada}
\altaffiltext{37}{University of Vienna, Institute for Astronomy, T\"urkenschanzstrasse 17, A-1180 Vienna, Austria}
\altaffiltext{38}{NASA Ames Research Center, Moffett Field, CA 94035, USA}
\altaffiltext{39}{Dept. of Astronomy and Astrophysics, University of Toronto, 50 St George Street, Toronto, ON M5S 3H4, Canada}
\altaffiltext{40}{Harvard-Smithsonian Center for Astrophysics, 60 Garden Street, Cambridge, MA 02138, USA}

\setcounter{footnote}{40}

\begin{abstract}
We report on both high-precision photometry from the \textit{MOST} space telescope and ground-based spectroscopy of the triple system $\delta$ Ori A consisting of a binary O9.5II+early-B (Aa1 and Aa2) with $P=$ 5.7d, and a more distant tertiary (O9 IV $P >400$ yrs). This data was collected in concert with X-ray spectroscopy from the Chandra X-ray Observatory.   Thanks to continuous coverage for 3 weeks, the \textit{MOST} light curve reveals clear eclipses between Aa1 and Aa2 for the first time in non-phased data. From the spectroscopy we have a well constrained radial velocity curve of Aa1. While we are unable to recover radial velocity variations of the secondary star, we are able to constrain several fundamental parameters of this system and determine an approximate mass of the primary using apsidal motion. We also detected second order modulations at 12 separate frequencies with spacings indicative of tidally influenced oscillations. These spacings have never been seen in a massive binary, making this system one of only a handful of such binaries which show evidence for tidally induced pulsations. 
\end{abstract}


\section{Introduction}

Massive stars are rare in the Universe, due both to the low--mass favored IMF \citep{sal55} and their short life times. However, their high luminosities and mass--loss rates make them  important to the ecology of the Universe. Underpinning this understanding is the observational determination of intrinsic stellar parameters for the most massive stars. However, this is not a simple proposition; except in rare cases where the distance is known reliably, it requires a detached, non-interacting binary system. We are aided by the fact that the binary fraction of O stars is nearly unity at birth, though this degrades some as the stars evolve (e.g. mergers and dynamical interactions: \citealt{sana14}). Still, there are only $\approx$ 50 O-star binaries in the Milky Way and Magellanic Clouds which meet these criteria and for which individual stellar characteristics have been established \citep{gies12}. This number may increase in the future owing to advances in interferometry, enabling one to determine stellar masses in systems such as HD 150136 (O3-3.5 V((f*))+O5.5-6 V((f))+O6.5-7 V((f)), 
\citealt{sana13}, \citealt{sanch13}). Right now, the small number of systems spanning the O spectral type makes calibrating models difficult, but an excellent effort to determine stellar parameters as a function of spectral type was made by \cite{mart09}.

Unfortunately, the problem of determining fundamental parameters becomes significantly worse for evolved O stars. It is extremely important to understand what effect leaving the main sequence has on the fundamental parameters of the star. However, there are currently only three such systems for which well-constrained masses have been derived, through the use of long baseline interferometry: $\zeta$ Ori A (O9.2 Ibvar Nwk+O9.5 II-III(n), \citealt{sota11}, \citealt{sota14}, \citealt{hum13}), HD 193322 (O9.5Vnn+O8.5III+B2.5V, \citealt{tenb11}) and MY Ser (O7.5 III + O9.5 III+(O9.5-B0)III–I, \citealt{iban13}). While increasing the total number of massive star systems with well known parameters is worthwhile, it is difficult as the number of candidate O stars is quite small. A much more realistic goal is finding a non-interacting example system with minimal mass loss which can be used as a template for understanding the evolution of O stars. Mintaka ($\delta$ Ori), the westernmost belt star of Orion, by virtue of its proximity ($d \approx 380 $ pc, \citealt{cab08}) and brightness ($V=2.41$, \citealt{mor78}), as well as encompassing the detached binary system Aa, is an excellent choice. 

The $\delta$ Ori system has been known to contain multiple components for well over 100 years \citep{hart04}. In fact, it is currently known to have at least 5 stellar components (a full diagrammed layout of this system and its relevant parameters can be found in Fig. 1 of \cite{harv02}). The embedded triple system, especially the short period binary $\delta$ Ori Aa, has received a significant amount of attention in the last 15 years. An artist's rendering of the orbits and lay out of the $\delta$ Ori A can be found in Fig 1. of \cite{tomer}. \citet[][hereafter H02]{harv02}, used spectra obtained with the International Ultraviolet Explorer to determine a putative secondary radial velocity curve. This, in combination with Hipparcos photometry, led to a controversial maximum mass for the O-star of 11.2 $M_{\odot}$, less than half the expected value based on its spectral type (see \citealt{mart09}). The system was revisited by \cite[][hereafter M10]{may10}, in an attempt to understand this anomalous result. They worked under the assumption that since the tertiary component contributed $\approx$ 25\% of the total light (\citealt{per97}, \citealt{horch01}, \citealt{mas09}, \citealt{maiz10}, \citealt{tok14}), significantly more than the secondary, its presence might be able to explain the apparent discrepancy.  They found that motion of the primary in combination with a wide tertiary  component with spectral lines broader than that of the primary gave the false impression of orbital motion of the secondary.  While M10 did improve the known orbital elements, the actual masses and radii are still poorly constrained. This is the result of a relatively weak secondary component as well as relatively poor light curve coverage and precision.

This paper is part of a series which will attempt to characterize as rigorously as possible the triple system $\delta$ Ori A, focusing on the O9.5II primary Aa1. This includes analysis of the $X$--ray spectrum by \citet[][hereafter Paper I]{mike} and $X$--ray variability by \citet[][hereafter Paper II]{joy}, as well as spectral modeling by \citet[][hereafter Paper IV]{tomer}. In this paper we focus primarily on the optical variability. This includes an analysis of high--precision space--based photometry obtained with the \textit{MOST} Space Telescope as well as simultaneous spectroscopic observations, described in detail in Section 2.  In Section 3, we discuss the spectroscopic variability, including a derivation of the primary radial velocity (RV) curve. In Section 4, we use this RV curve in conjunction with the \textit{MOST} light curve to provide three possible  binary fits at varying values of the primary mass. In Section 5, we provide a fit to the primary mass using apsidal motion confirming our results from section 4. Finally, we discuss and characterize previously unknown photometric variability within this system and report frequency and period spacings consistent with non-linear interactions between tidally excited and stellar g-mode oscillations. While we are still are unable to disentangle the secondary spectrum and its kinematics, we provide a more complete model than has been achieved up to this point.

\section{Observations}

\subsection{High Precision Photometry from {\it MOST}}

Our optical photometry was obtained with the {\it MOST} microsatellite that houses a 15-cm Maksutov telescope through a custom broad-band filter covering 3500--7500 \AA. The sun-synchronous polar orbit has a period of 101.4 minutes ($f=14.20 \ {\rm d}^{-1}$), which enables uninterrupted observations for up to eight weeks for targets in the continuous viewing zone. A pre-launch summary of the mission is given by \cite{most1}.


\begin{figure}[ht]
	\centering
	\includegraphics[scale=0.7]{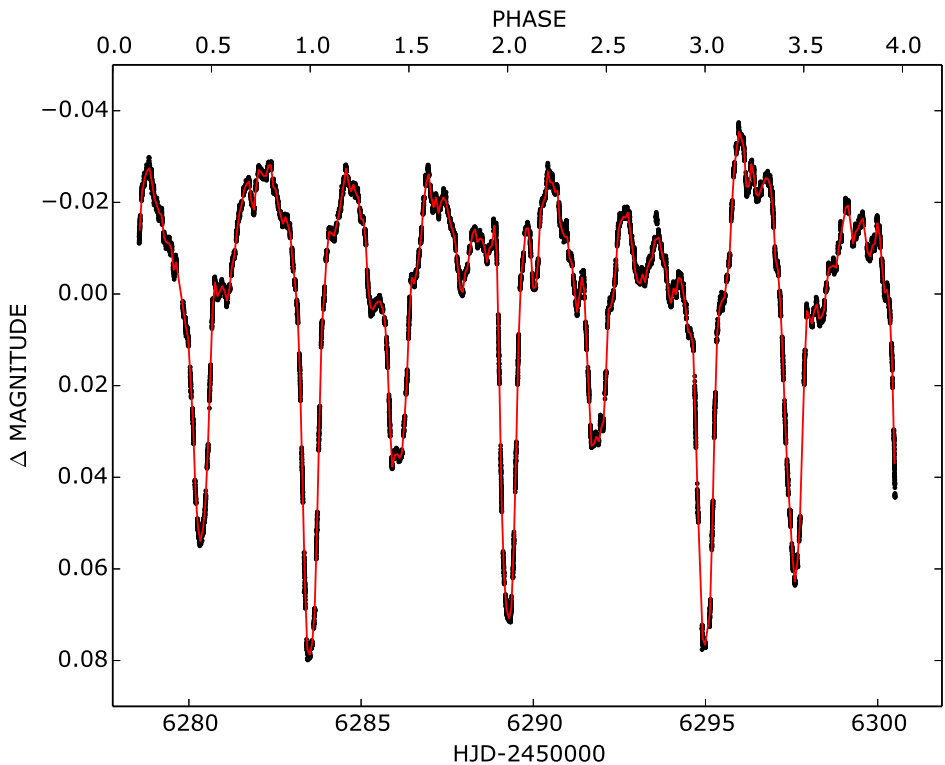}
	\caption{ {\it MOST} light curve (black) of $\delta$ Ori taken in mid December 2012--early January 2013. Error bars are smaller than the size of the points. The red line is the combination of the binary fit (see Section 4) and that of the secondary variations (see Section 6).}
	\label{fig:full-lc}
\end{figure}

$\delta$ Ori A was observed for roughly half of each {\it MOST} orbit from  Dec. 2012 through 7 Jan. 2013 with the Fabry-mode (see Fig. \ref{fig:full-lc}) at a cadence of 40.4 s. These data were then extracted using the technique of \cite{most3}, and show a point-to-point standard deviation of $\sim 0.5$ mmag. 

\subsection{Optical Spectroscopy}

We initiated a professional-amateur campaign in order to obtain a large number of high-quality optical spectra simultaneous with our {\it MOST} and {\it Chandra} campaigns. This resulted in more than 300 moderate resolution ($R \gtrsim 10000$) spectra obtained over the 3 week period. The spectra were reduced by standard techniques utilizing bias, dark, and flat field images. Wavelength calibration was accomplished through comparison with emission-lamp spectra taken before and/or after the object spectra. Our main goal was originally to monitor the H$\alpha$ line for wind variability  and correlate any such emission with the X-ray spectra (Paper I, Paper II) while simultaneously obtaining a measure of the orbital motion from the \ion{He}{1} 6678 transition. However, no significant H$\alpha$ variability was seen, except for the radial velocity changes related to the orbital motion of Aa1. Furthermore, H$\alpha$ could be diluted by wind emission, making RV measurements questionable. Due to both the telluric absorption lines around H$\alpha$ and the potential for wind emission, we focused on the He I 6678 transition for our radial velocity measurements. Details of the telescopes and spectrographs used are given in Table \ref{table:speclog}. A typical S/N for any observation is $\approx$ 100--150.

\section{The Spectroscopic Orbit}

The spectroscopic treatment of $\delta$ Ori A from M10 and H02 are not in agreement. The M10 analysis assumed a tertiary contribution that was subtracted off, and left a single-lined binary, while the H02 analysis found evidence of the tertiary after iterating to remove first the primary and then the secondary spectrum from the combined one. The resulting masses are highly discrepant, with M10 assuming a standard mass based on spectral type of $M_{\rm primary} = 25 M_{\odot} $, while H02 found $M_{\rm primary} = 11.2 M_{\odot}$. Treatment of the tertiary may lead to severe errors in the inferred stellar parameters.

Paper IV presents a quantitative photospheric and wind model for each of the three stars in the $\delta$ Ori A system. The results of \cite{mas09} and \cite{tok14} show that the tertiary orbits the binary Aa very slowly with a period of several hundred years, although the uncertainties in the tertiary's orbit are very large. As the period is so long, we expect no measurable kinematic motion in the tertiary component during one month of observations. Therefore, if we take the model of
the tertiary spectrum from the models of Paper IV, then we can subtract its contribution a priori from the spectra in order to best constrain the binary properties of Aa1,2. This is similar to what was done by M10. If the tertiary is not taken into account, then our results are in agreement with H02. However, the optical spectrum consists of $\approx$ 25 \% contribution from the tertiary (\citealt{per97}, \citealt{horch01}, \citealt{mas09}), so even with our S/N $\approx$ 100--150, this contribution should always be seen in our spectra.

\begin{figure}[htp]
	\centering
	\includegraphics[scale=0.6,angle=90]{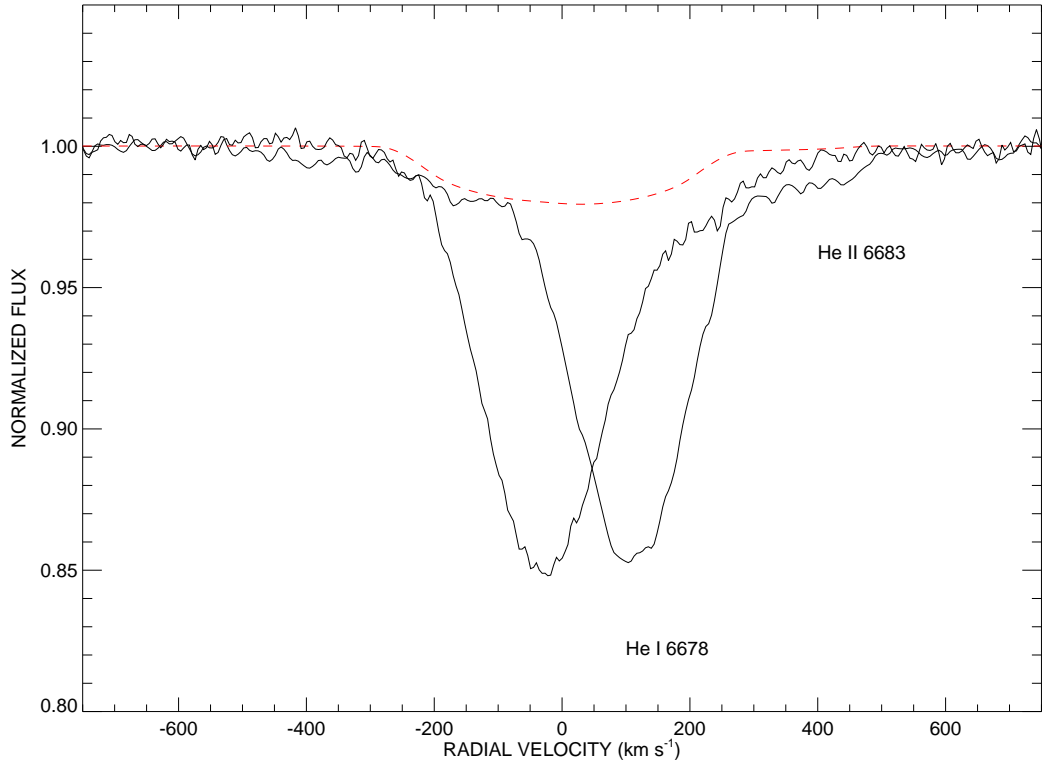}
	\caption{ Two observed He I 6678 profiles (from HJD 2456279.99 showing the negative velocity extremum, and 2456281.75 showing the positive velocity extremum). The He II 6683 absorption shown by a dashed red line represents
a weighted non-moving spectrum of the tertiary calculated by Paper IV.}
	\label{tert_oplot}
\end{figure}

We convolved the calculated spectrum of the tertiary to the resolution of our data, and can immediately see that it is the likely source of excess absorption. In Figure \ref{tert_oplot}, we show two example profiles of the He I 6678 line (with large red and blue shifts from the primary's orbital motion) along with the weighted contribution of the tertiary's spectrum calculated by Paper IV. For the redshifted spectrum, we see that the red edge of the profile has an excess of absorption that matches the theoretical spectrum of the tertiary quite well. On the blue edge of the profiles, blending with He II 6683 becomes more problematic. While the tertiary seems to reproduce some of the absorption, most of blue excess comes from the primary star's He II line. 


After we subtracted the weighted contribution from the tertiary star, we cross correlated our observations against the model of the primary calculated by Paper IV. From these velocities we calculated a single-lined orbit with the period derived by M10, as the length of our data sets was insufficient for determining a more precise period. Most velocities had uncertainties on the order of $\pm2$ km s$^{-1}$ from this determination. Data for all the velocities obtained are given in Table 4.

We examined the residuals after removing the primary star's spectrum (both through shift-and-add and from model subtraction using the spectrum calculated by Paper IV). We found no residuals due to the secondary in the data at this point. This is consistent with the observational results of M10 as well as the results of the modeling of Paper IV. In particular, the modeling results predict that only $5.8\%$ of the light in this wavelength region comes from the secondary, so the noise associated with the weak features would dominate, as the individual spectra have a signal-to-noise on the order 100--150 on average. In order to reliably see the secondary with a S/N of 50, we suspect that the combined spectrum of the three component stars needs to have a S/N $\gtrsim$ 1000. 

\section{A Full Binary Model}

A complete solution for a binary system requires a substantial amount of information. In general, it requires an eclipsing light curve  and two radial velocity curves. Despite being studied numerous times, no one has been able to reliably determine all three of these quantities for $\delta$ Ori Aa. This is primarily due to the faintness of the secondary with respect to both the primary and the tertiary. Despite the lack of a secondary RV curve, data have two significant advantages over the previous works of H02 and M10: high precision  photometry along with simultaneous spectroscopy. Since there is known apsidal motion in this system (see Section 5) having all photometry and  spectra taken all within a month timespan mitigates this issue. 

\subsection{Initial Setup and Modeling}

For simulation and fitting of the $\delta$ Ori system we used the alpha release of PHOEBE 2.0 (\citealt{pieter},  \citealt{andrej}), an engine capable of simulating stars, binaries, and concurrent observational data. Unlike the original PHOEBE, this is not built directly on the work \citep{wd71} and has been rewritten to take into account many second-order effects. This is especially relevant when dealing with spaced-based data.

Before this system was fit, some initial processing was required. As shown in Fig.\ \ref{fig:full-lc}, the light curve of $\delta$ Ori is composed of an eclipsing light curve with second order variations superimposed. These underlying variations cannot be easily or completely removed (see Section 6) so we must lessen their effects in order to get a more accurate representation of the binary variability. This can be done by first phasing on the binary period, which requires a precise ephemeris for the system. Since our data span only 3 weeks (about 3.5 orbits), the period determined by M10 is much more precise than what we are able to obtain, so we adopt their value $P = 5.732436 \pm 0.000015$ d. For $T_{0, {\rm min}}$ and $T_{0, {\rm periastron}} $  we adopted our own derived values. For $T_{0, {\rm periastron}} $ we determined a value from fits to the radial velocity curve of $2456295.674 \pm 0.062$ HJD. For $T_{0, {\rm min}}$ we are actually able to obtain a value from O-C calculations of $2456277.7897 \pm 0.0235$ HJD. We should note that these numbers  offer no specific improvement in the associated error over those derived in  M10, and in the case of $T_{0, {\rm min}}$ the error is slightly worse. However, unlike M10  our method for determining $T_{0, {\rm min}}$ is independent of the binary fit. This is important as second order stellar effects (unknown in previous works) make determinations from the binary fit less reliable. In addition, apsidal motion is significant in this system, and we did not want it to affect our results.

\begin{figure}[htp]
	\centering
	\includegraphics[scale=0.7]{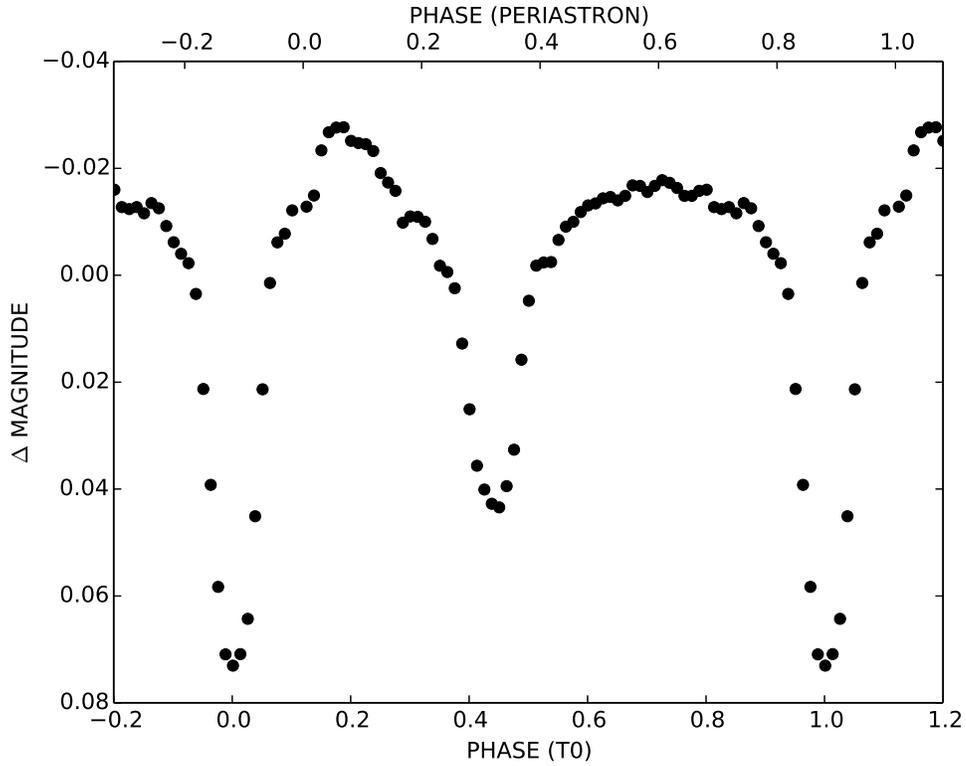}
	\caption{ The \textit{MOST} light curve of $\delta$ Ori phased on the orbital period. It is then binned to 0.0125 in phase to smooth out remaining secondary variations. The bottom x-axis is phased to time of minimum light and the top is phased to periastron. }
	\label{phlc}
\end{figure}

Using the revised ephemeris, the data were phased and then binned to provide a mild smoothing, which yields a template for the binary variations. In this template, the error is much smaller than the size of the points. However, incomplete removal of the secondary variations leads to scatter at the milli-mag level (see Fig. \ref{phlc}). 


In addition, modeling of stars and binary systems requires the presence of realistic atmosphere models. Unfortunately, the Kurucz atmosphere tables are the only ones which have been incorporated into PHOEBE 2.0 at this stage and they are insufficient for O stars. Therefore, our best option is to use blackbody atmosphere models. In some cases this would not be adequate, but for O stars we sample the Rayleigh--Jeans tail at optical wavelengths, making the differences from a blackbody minimal. In our specific case, a comparison with non-LTE models calculated for the components of the system indicates a deviation of $\approx 5$\% in the integrated flux across the \textit{MOST} bandpass  (Paper IV). Therefore, this assumption will not significantly affect our solution as all three stars have similar temperatures and deviations from a blackbody. 


\begin{figure}[htp]
	\centering
	\includegraphics[scale=0.7]{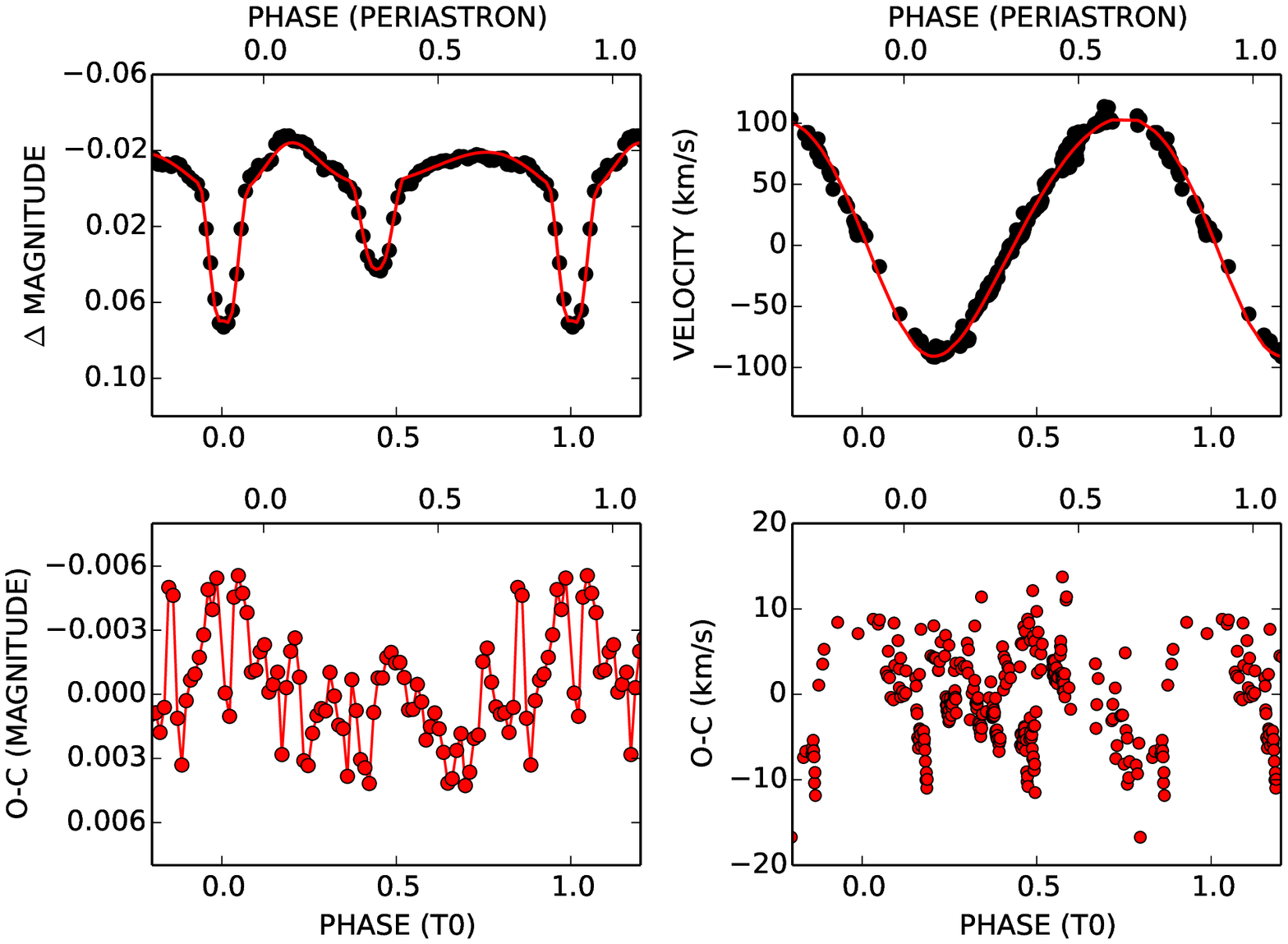}
	\caption{The top two panels show the binned \textit{MOST} light curve (left) and the radial velocity curve (right) overlaid with the low mass PHOEBE 2.0 fit (see Table \ref{table:bin-fit}). In the bottom panels are the O-C residuals for the corresponding light and radial velocity curves. The intrinsic stellar variations that we are unable to completely remove are the main source of error. Only one model is plotted as the differences are nearly impossible to distinguish at this scale.}
	\label{fig:bin-fit}
\end{figure}

\subsection{Fitting and Solution}

PHOEBE 2.0 has several built--in fitting routines, but for the purposes of this project we decided to use the Monte Carlo Markov Chain method (MCMC). The specific implementation used by PHOEBE 2.0 is emcee \citep{emcee}, an affine invariant MCMC ensemble sampler based on \cite{good10}. While this routine requires a good initial fit to converge to a solution in a reasonable number of iterations, it gives the most comprehensive analysis of both the fit and the parameter correlations. MCMC also has become a standard in the fitting of photometric time series (e.g. \citealt{greg05} and \citealt{crol06}). For a more general overview of this technique, see \cite{gilks96}. 

However, this fitting method is not particularly useful without a full complement of constraints. Since we have no secondary RV curve, we must put an extra constraint on the mass. We know the spectral type of the primary star quite well and can make a reasonable assumption that the mass is $25 \pm 4 M_{\odot}$ \citep{mart09}, further supported by non-LTE modeling of the system by Paper IV. Unfortunately, as there are several viable sets of parameters, the fit was unable to converge to a  single minimum. Therefore, we divided this mass range further into three separate ranges: low mass primary (21-24 $M_{\odot}$; LM) , medium mass primary (24-27 $M_{\odot}$; MM), and high mass primary (27-29 $M_{\odot}$; HM). These mass ranges were chosen based on previous results, as the system would converge to one of three possible solutions, with one falling in each bin. 

For each possible solution we iterated at least 800 times. While this may seem insufficient to converge to a true solution, we were not starting from a random point in the parameter space. Parameters were adjusted so that the overall shape and amplitude were consistent with the observed light curve and RV curve before beginning MCMC. Indeed, after about 200 iterations the log probability function defining our goodness of fit is virtually constant. 

\begin{figure}[htp]
	\centering
	\includegraphics[scale=0.6]{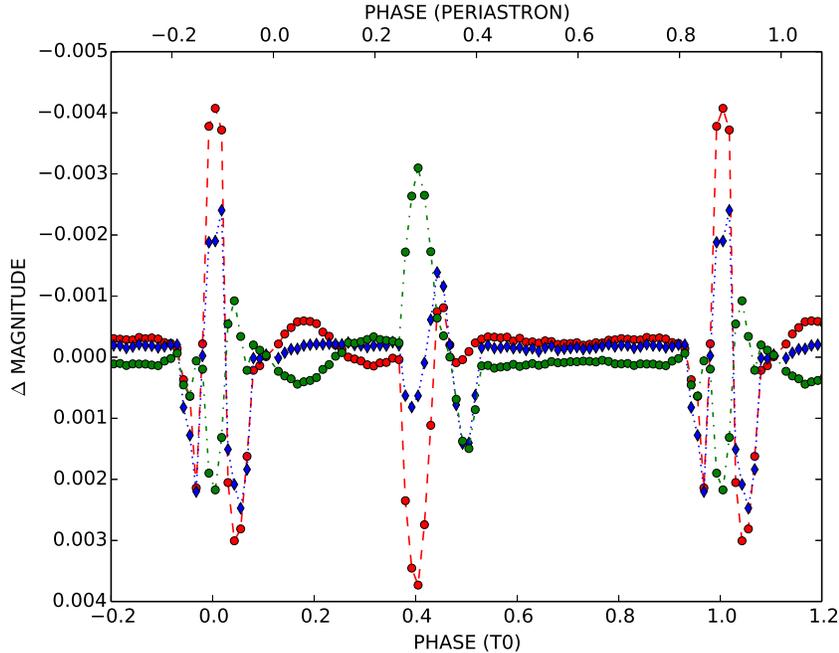}
	\caption{ Shown is a comparison of the residuals of each fit: LM-MM (red), LM-HM (blue), and MM-HM (green). Each fit has a standard deviation of around 2.5 mmags. They are very nearly identical to each other in all places except in the eclipses.}
	\label{fig:res-comp}
\end{figure}

A representation of the quality of these fits for the LM model is shown in  Fig. \ref{fig:bin-fit} with the upper panels showing the PHOEBE fit to the light curve (left) and RV curve (right). The lower panels show the corresponding residuals.  While there are differences between the three models, they are impossible to distinguish on this scale and so only the LM model is shown. The only motivation for this model over the others is that its primary mass agrees best with what is derived from apsidal motion (see Section 5). One can see a comparison of the light curve residuals of each of these models in Fig. \ref{fig:res-comp}.  This figure demonstrates the differences inherent in the residuals of each model. These differences are small but still significant, especially within the eclipses. However, it is impossible to determine quantitatively which solution is best as the standard deviation of the residuals for each fit are the same to within $\approx 2 \times 10^{-5}$. It should also be noted that PHOEBE 2.0 is still undergoing rigorous testing, so while our models seem reasonable, we also simulated our solutions with the original PHOEBE \citep{phoebe1} to ensure that the results were consistent.


Table \ref{table:bin-fit} shows the complete solutions for each model as well as comparisons to previous solutions by H02 and M10. While the spurious detection of the secondary RV biases the H02 results, there is good agreement between our work and that of M10. One point to note is the improvement made to the inclination angle. While each of the previous works had fits at substantially different inclinations from 67$\degree$ - 81$\degree$, our work shows that this inclination is quite firmly centered at 76$\degree \pm 4 \degree$ regardless of the fit. In addition all of the orbital parameters are well constrained. It is very difficult to say if the high, low, or medium mass fit is the best. The probability function is virtually identical for the three fits, and the only noticeable differences are those for the masses and radii of the primary and secondary. In both the RV and light curve solutions the residuals are due more to the accuracy of the points than in any differences in the model (see Fig.  \ref{fig:bin-fit} and Fig. \ref{fig:res-comp}). Therefore, we must allow for a family of solutions until we are able to retrieve a secondary RV curve. With no clear way to select a best fit model, we adopt the LM fit throughout this paper when we need to apply a model.

\section{Apsidal Motion}

In close binaries such as $\delta$ Ori Aa, the deformations of the stars in the elliptical orbit cause the longitude of periastron ($\omega$) to precess with time, which is referred to as apsidal motion. The amount of this precession depends on the mass, internal structure of the stars, and the orbital eccentricity. \cite{ben02} and \cite{fer13} made assumptions for structure constants of massive stars and then used the measurements of apsidal motion to calculate stellar masses.

\begin{figure}[htp]
	\centering
	\includegraphics[scale=0.6,angle=90]{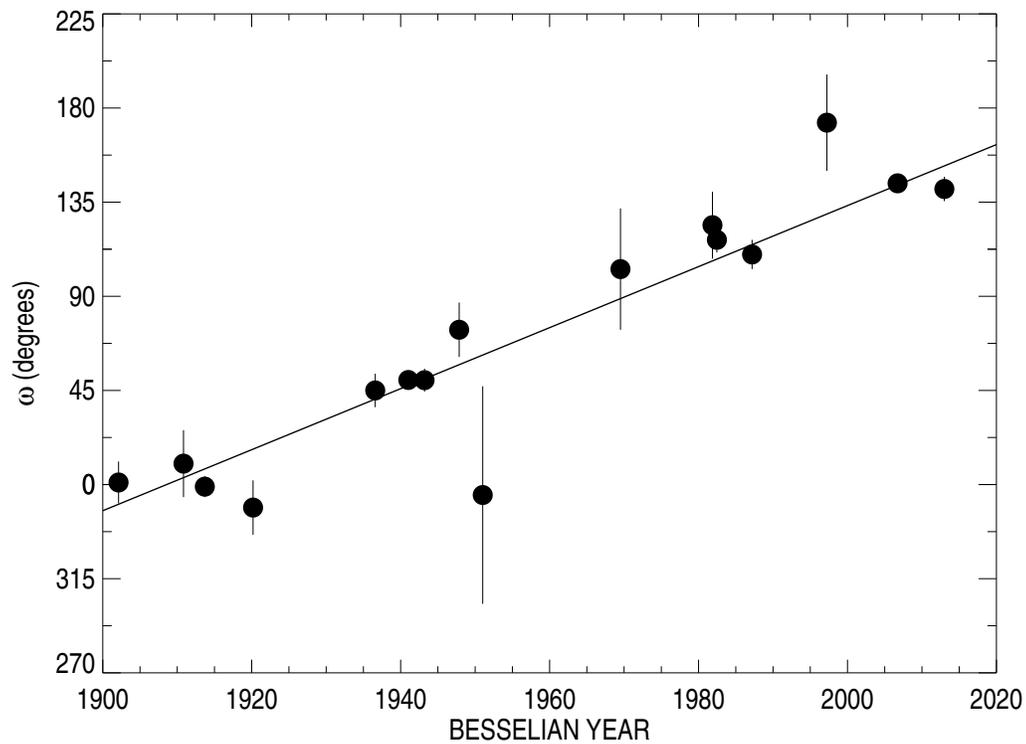}
	\caption{ Apsidal motion of $\delta$ Ori. A weighted linear fit to the published values (\citealt{harv87}, H02, M10) yields a measurement of $\dot{\omega} = (1.45 \pm 0.04)^{\circ}{\rm yr}^{-1}.$}
	\label{apsidal}
\end{figure}

$\delta$ Ori A has been studied extensively for more than a century and shows signs of significant apsidal motion. The work of \cite{harv87} compiled all determinations of the spectroscopic orbit. This is combined with more recent studies of H02 and M10 along with our own work to retrieve orbital information spanning over 100 years. We use these derived orbital elements to calculate the apsidal motion, $\dot{\omega}$, using a weighted linear least-squares fit to $\omega$ as a function of time. The results shown in Figure \ref{apsidal} present a reasonable fit to the measurements and provide a value of $\dot{\omega} = (1.45 \pm 0.04)^{\circ}{\rm yr}^{-1}.$


\begin{figure}[htp]
	\centering
	\includegraphics[scale=0.6, angle=270]{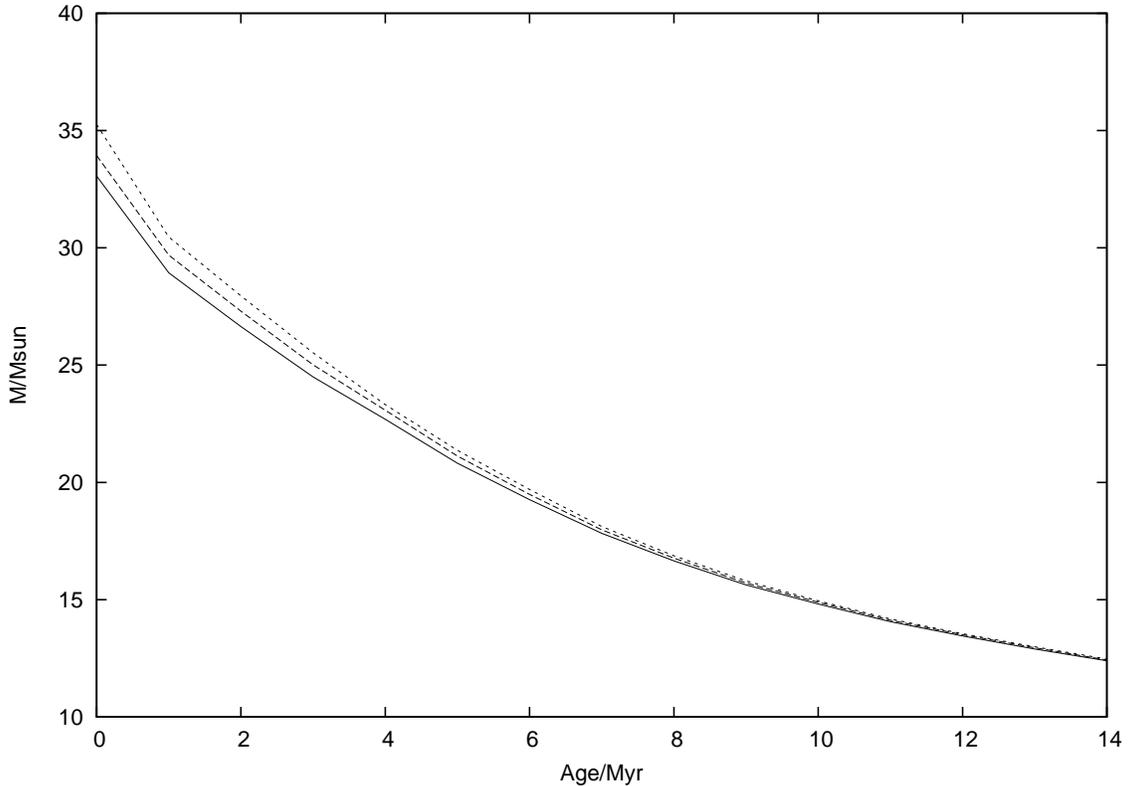}
	\caption{ Derived mass as a function of age from the apsidal motion and assumed spectroscopic parameters of $\delta$ Ori. \cite{cab08}  estimate the age of the cluster containing $\delta$ Ori, to be 5 Myr, corresponding to a mass of $\approx 22 M_\odot$ consistent with our low mass model. The different lines represent the model determinations with the extremes of the apsidal motion from the errors (Fig.~\ref{apsidal}).}
	\label{apsmass}
\end{figure}

With this measurement of the apsidal advance, we can make some assumptions about the internal structure constants of the primary star. This relies heavily on the age of the system. The age of $\delta$ Ori is likely best constrained by the study of \cite{cab08} who find $\delta$ Ori Aa1 to be the brightest star in a small cluster. Their study involved deep imaging to understand low-mass stars, and their color-magnitude diagrams imply an age of $\approx 5$ Myr. In order to continue with the calculation we adopt our orbital parameters from \S 4.1. Combining these and the models of \cite{ben02}, we obtained the results shown in Fig. \ref{apsmass}, which give the stellar mass as a function of age for our derived apsidal motion. With an age of 5 Myr, we see that the derived mass of the primary is $\approx 22 M_\odot$.  This value agrees with the spectral--type/mass \citep{mart09} and our low mass model (Table \ref{table:bin-fit}).


\section{Photometric Time-Series Analysis}

It is clear from Fig. \ref{fig:full-lc} that the dominant signal in the light curve comes from binarity. It is also apparent from the non-phase-locked variations that additional signals are present in $\delta$ Ori A. The presence of such variation is not completely unexpected as \cite{khol06} reported spectral variability in this system. However, the variation timescales seen in Fig.\ref{fig:full-lc} are much longer than the  $\approx$ 4h periodicities they report. 


\begin{figure}[ht]
	\centering
	\includegraphics[scale=0.6]{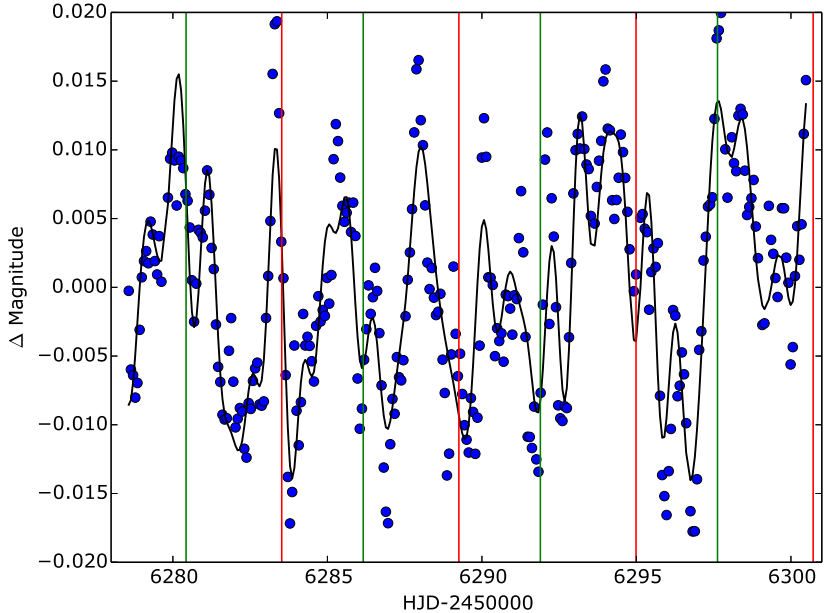}
	\caption{ The {\it MOST} light curve with the binary signal removed. The residuals show the presence of obvious secondary variations in the light curve. The black line shows the Period04 fit to the data. The lines show the location of the primary (red) and secondary (green) eclipses.}
	\label{fig:resids-lc}
\end{figure}
Before addressing this variation, we first bin our light curve with the \textit{MOST} orbital period (0.074 d). This allows us to mitigate the problem of light scattered into the \textit{MOST} optics (mainly due to the bright Earth). As a consequence, we ignore variability on time scales shorter than a single orbit. Once this is done, we must remove the eclipsing binary signal. This variability has already been analyzed in the previous section, and hinders our ability to study the additional variations present. 

There are two options available for removal of the binary signal. We could use the binary template derived earlier (see Fig. \ref{phlc}) or the LM model fit to this template. The decision of which  binary light curve to use is not obvious, so the analysis was done using both subtractions. The results are largely the same, and we chose the LM model fit subtraction as the amplitude of the significant peaks was higher and there were slightly more significant frequencies. This choice is supported by the agreement between the subsequent analysis and the full light curve (see Fig.~\ref{fig:full-lc}, red line).
The residuals are shown in Fig.~\ref{fig:resids-lc} (blue dots).    


\begin{figure}[ht]
	\centering
	\includegraphics[scale=0.6]{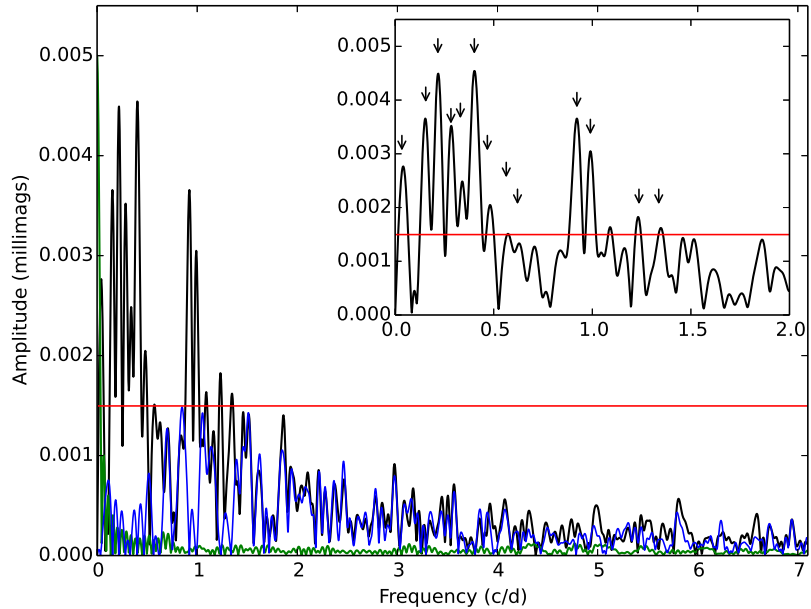}
	\caption{Fourier transform of the binary subtracted residuals of $\delta$ Ori A (black).The spectral window is plotted in green and shows no significant aliasing. The Fourier transform after pre--whitening is shown in blue. The red line at 1.49 millimag represents the $4\sigma$ significance.   The inset is a zoom in of the Fourier transform from 0--2 c/d. Pre--whitened frequencies are labeled with arrows (see Section 6). Note that some frequencies are not immediately apparent without removal of larger amplitude peaks.}
	\label{fig:del-ft}
\end{figure}

A common method for finding and classifying periodic signals is known as pre-whitening. In this method peaks are detected in the Fourier transform and then a sinusoid with the corresponding period, amplitude, and phase is subtracted from the light curve. This is repeated, and a combination of sinusoids representing each frequency are fitted and removed until there are no more significant peaks present. This method only works if the signals are sinusoidal (or mostly sinusoidal) and in the case of $\delta$ Ori A, this is not obvious. 

The Fourier transform of the $\delta$ Ori A light curve is shown in Fig.~\ref{fig:del-ft} (black) using Period04 \citep{p04}.  We apply pre--whitening to the transform and the end product is shown in blue. In order to determine which frequencies to include as significant, we use an empirically determined amplitude-to-noise ratio of 4 \citep{breg93}.  There are  12 frequencies (black arrows) above this significance threshold $4\sigma = 0.00149$ (red line in Fig. \ref{fig:del-ft}). The parameters associated with each frequency are given in Table \ref{table:ft-peaks}. The fit of these frequencies to the residuals, which is shown in Fig. \ref{fig:resids-lc} (black line), does a reasonable job of fitting the variations. While \cite{khol06} attribute the variability they find to non-radial pulsations, we see no significant variability in the spectrum around 4 hrs (6 c/d). 


\begin{figure}[ht]
	\centering
	\includegraphics[scale=0.6]{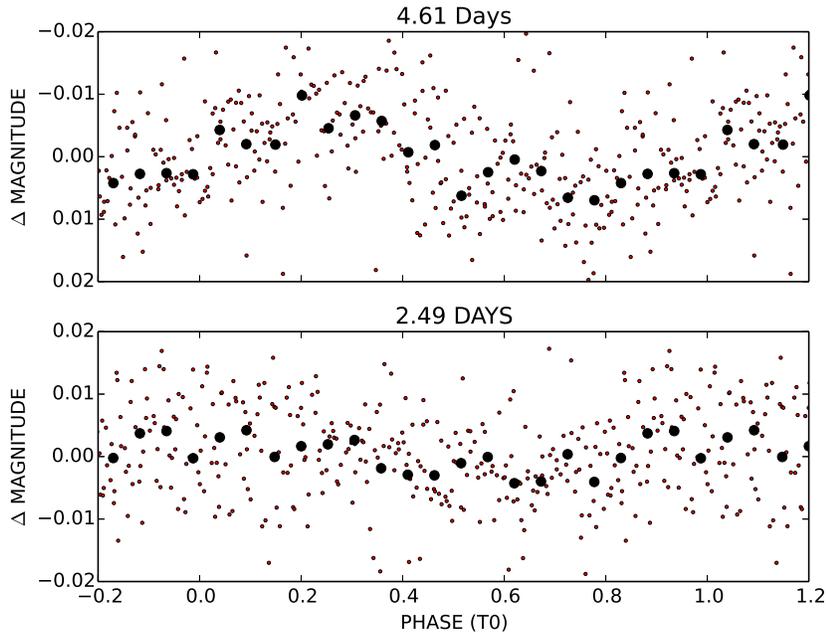}
	\caption{ Both plots show the phased residuals (small red points) overlaid with the binned (large black points) data to emphasise the shape. The top plot shows the largest amplitude peak (4.61 d). In the bottom plot this signal has been removed and the data is phased on the next highest frequency (2.49 d). }
	\label{fig:period-lcs}
\end{figure}

One possible explanation of extra periods is the presence of rotational modulation through starspots. While this is rarely considered in massive stars, the models of \cite{cant11} lend some credibility to this idea. Moreover, the work of \cite{tahina14}, \cite{blomme11}, and \cite{hen14} show evidence for the occurrence of this phenomenon among O stars. If rotational modulation is the root cause of the variation, then the largest frequencies would occur near the rotation period of the star. The vast majority of the remaining peaks could be explained as spurious signals due to the variable nature of star spots (e.g. lifetimes and number) as explored extensively by \cite{tahina14} for the O7.5 III(n)((f)) star $\xi$ Per. In our case the period of the largest amplitude signal is 4.61 d. If we assume this this is the rotation period of $\delta$ Ori Aa1, then combined with the $v \sin i \approx 130 \ {\rm km s}^{-1}$ (Paper IV) and the assumption of alignment between the orbital and rotation axis, this yields a radius of $\approx 12R_\odot$ for the primary Aa1. This is smaller than the value found in our models (see Tab. \ref{fig:bin-fit}) and closer to that of a luminosity class III star \citep{mart09} as opposed to class II which $\delta$ Ori Aa1 is known to be. However, it is unclear what the uncertainty in radius would be in the models of \cite{mart09}  and so this peak remains roughly consistent with rotational modulation. Indeed, the phase folded light curves of the two strongest frequencies (see Fig. \ref{fig:period-lcs}) show evidence of non-sinusoidal behavior. This could be a consequence of a highly non-sinusoidal signal such as rotational modulation.

\begin{figure}[ht]
	\centering
	\includegraphics[scale=0.6]{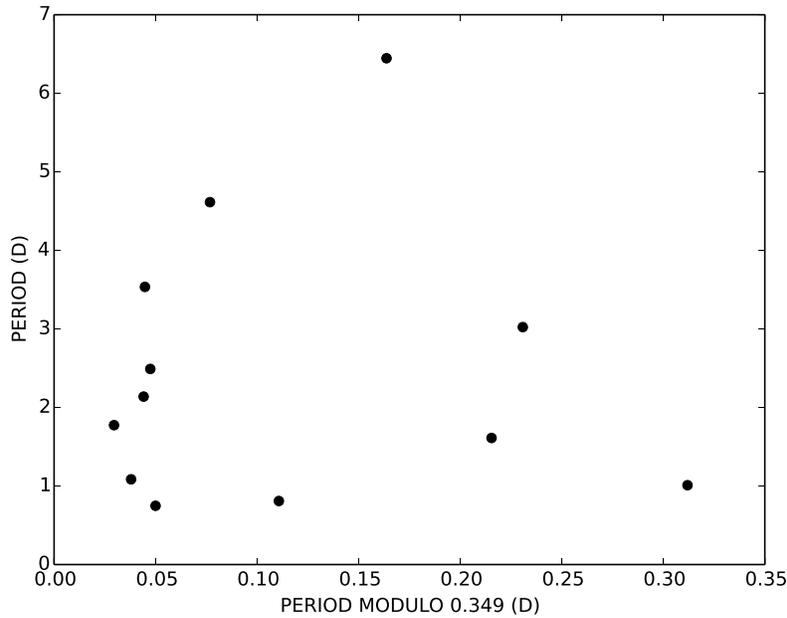}
	\caption{ Echelle diagram phase--folded over a period of (0.349 d). Note the line of periods which fall around 0.05 indicating that they share a common period spacing.}
	\label{fig:echelle}
\end{figure}

While this hypothesis is plausible, similar variations can also be caused by pulsations.  While rotational modulation would result in a forest of peaks that show no clear relationship to each other \citep{tahina14}, pulsations should show evidence of coherent spacings. All the periodicities we observe are several hours to days long, therefore they would fall in the g-mode regime for all three massive stars seen in this system \citep{pam14}.  These modes have the unique quality that when they have the same $l$ number they are equally spaced in period. Further explanation can be found in any asteroseismology review (e.g. \citealt{aerts10}. Now, we can check for period spacings by creating an echelle diagram where the period of each variation is phased modulo a given spacing and plotted. Equally spaced periods will lie roughly along a vertical line. The echelle diagram in Fig. \ref{fig:echelle} shows that at least half of the frequencies share a common period spacing. The results of this analysis provide evidence for pulsations, but this evidence is no more convincing than the idea of rotational modulation. However, there also appear spacings in the frequency domain. Curiously, 10 of the 13 frequencies show a spacing at two or 3 times the orbital frequency (see Tab. \ref{table:spacings}). This spacing  suggests that there are tidal interactions associated with these signals. This is not unprecented as  the work of \cite{pal14}, \cite{puls} and \cite{pal13} also show some evidence of effects caused by tidal forces in massive binaries.

This variability, though, is unlikely to arise from tidally excited oscillations, which occur at integer harmonics of the orbital frequency (\citealt{full12}, \citealt{burk12}, \citealt{welsh11}), in contrast to the oscillations we observe. However, frequency spacing at multiples of the orbital frequency has been seen in several other close binary stars (\citealt{hamb13}, \citealt{bork14}, \citealt{mac14}), and indicates that the pulsations are tidally ``influenced." Similar to the papers cited above we speculate that the frequency spacing at multiples of the orbital frequency may arise from non-linear interactions between stellar oscillation modes and tidally excited oscillations. Moreover, we interpret the frequency spacing as evidence that the observed variability is generated by stellar oscillations rather than by magnetic activity or rotational modulation.

The main caveat to this hypothesis is that our frequency resolution is limited to 0.047 d$^{-1}$ due to the short span (3 weeks) of the observations. This means that there are several spacings that could have been deemed significant. For instance $F7-F11$ and $F2-F11$ would both, within error, fit the same spacing because of how close the two peaks are in frequency. When this happens, we choose the peak that resulted in a spacing closest to a multiple of the orbital frequency. Additionally, the two frequencies which do not appear in Tab. \ref{table:spacings}, $F4$ and $F5$, are actually split by the orbital frequency. However, since the uncertainty is half the actual spacing, we choose not to include it.  Despite these issues, this frequency spacing is pervasive, appearing strongly between nearly every significant peak. While there is still uncertainty in this claim, it can be lifted with the addition of longer time baseline observations. Another problem is that even if these are tidally ``influenced" oscillations, they still do not completely account for all the variation in this system. It is possible that these discrepancies could be due to amplitude modulation, though the ratio of period versus observation length makes this unlikely. It is also possible that there is also rotation modulation in this system from one of the components, just at a lower level than the pulsations. 

It is worth noting that the nature of the significant periodicities found in the
lowest frequency range of the \textit{MOST} data set is uncertain and could be a result of 'red noise'. Normally, this noise could be modeled and removed using Autoregressive Moving Average techniques. However, our data is not stationary which is a requirement for this type of modeling. Attempts at making the data stationary disrupt the relevant signals; not only in our data, but also in the synthetic data using combinations of purely sinusoidal signals. Therefore, it is impossible to determine unequivocally the root cause of this variation without the addition of longer baseline photometric data (see Section 8).

\section{Discussion}

The system $\delta $ Ori A remains an ideal candidate to define the intrinsic parameters of evolved O stars.  The fundamental parameters of the primary, $\delta$ Ori Aa1, appear consistent with models from Paper IV as well as the results from modeling of the apsidal motion. Our analysis removes ambiguity in the derived stellar parameters and makes the primary more consistent with stellar models, without having to appeal to non-conservative mass loss or mass transfer that might play a role in the evolution of  a close binary like $\delta$ Ori Aa.

Despite these consistencies, $\delta$ Ori Aa  has a rather larger eccentricity of $\approx 0.114$, which is unusual for a close O-star binary system with a moderate age of $\approx$ 5 Myrs. In such a situation, circularization of the orbit should already have occurred. We can check this using the procedure outlined in \cite{zahn77} with one key assumption: knowledge of the stellar structure constant $E_{2}$. This constant, which is dependent on the interior composition and structure of the star, requires extensive effort to calculate precisely, but can be approximated by the fractional core radius to the eighth power. We have no evolutionary models for $\delta$ Ori Aa so we have no way to calculate or estimate $E_{2}$ directly. We do know from \cite{zahn75} that $E_{2}$ tends to increase as a function of mass, so we will take the value of $E_{2} \approx 3.49 \times 10^{-6}$ which is that of a 15 $M_{\odot}$ luminosity class V star. We choose this value as this is the largest mass star for which a value of $E_{2}$ was derived. This is clearly an underestimate of $E_{2}$  which  is ameliorated slightly by the fact that $\delta$ Ori Aa is a slightly expanded luminosity class II star (implying a smaller fractional core radius). It is also clear based on the inverse dependence of the circularization time with $E_{2}$ that the value we obtain will be a maximum. A quick calculation reveals a circularization timescale of $\approx 1$Myr, which is well short of the age estimate that has been derived for this system. This result shows that our knowledge of this system is incomplete and could indicate that our assumed primary mass is incorrect. However, this discrepancy could also be caused by incorrect age calculations, strong interactions with the tertiary component, or issues with our knowledge of stellar structure in massive stars. At the same time, there is currently no evidence which strongly supports any of these hypotheses, so  we are confident in the assumptions made.
 
The most intriguing aspect of this system is the pronounced second order variations in its light curve.  Such variations have never been seen before in this system, and the frequency spacings seem suggestive of tidally induced pulsations.  If confirmed, this would be an important discovery as these modes have been seen rarely if ever in massive stars. This could lead to asteroseismic modeling and increase our understanding of the interiors of both massive stars in binaries, and massive stars as a whole, especially considering the extremely high fraction of these stars in binaries \citep{sana14}.
 
While we have made significant progress towards understanding this system, it is clear that we are not at the level we need to be. Our ultimate goal is for $\delta$ Ori to be a test for both evolutionary and structural models of massive stars. For this objective to be realized, we must determine fundamental parameters to high precision. For this reason it  is imperative that we identify the presence of the secondary component within the spectrum. In addition, we must improve our frequency resolution within the Fourier spectrum. This will require a long time-series of high-precision photometry. As a result, we could identify with confidence the source of the variation, which could possibly allow for asteroseismic modeling. A secondary consequence of this would be an improved binary light curve with significantly reduced scatter from which to model our system. This would also lead to improved values of fundamental parameters, and place constraints on the apsidal advance of the system.

\section{Future Work}
This analysis to understand $\delta$ Ori, in conjunction with Papers I, II, and IV, has highlighted the need to obtain the spectral characteristics and RV curve of $\delta$ Ori Aa2. We will use both the {\it Hubble Space Telescope} and {\it Gemini-North} to obtain high-resolution spectra of the binary at opposite quadratures while spatially separating the tertiary and observing its spectrum. These observations are scheduled for the upcoming observing season, and we also plan to obtain higher signal-to-noise optical spectroscopy from the Observatoire de Mont M\'{e}gantic. These observations will provide excellent constraints on the modeling of the system and reveal any unusual spectroscopic variability seen in the tertiary star. This will provide more constraints on the photometric variability seen with \textit{MOST} .

A better characterization of this variability as well as the binary parameters requires access to long time baseline photometry, which \textit{MOST} is not able to provide. The \textit{BRITE-Constellation} project, however, is designed to provide 6 months of continuous coverage in both blue and red filters \citep{brite}. \textit{BRITE-Constellation} began initial data acquisition of the Orion constellation in October of 2013. Since this was the initial data taken, the full 6 months of continuous coverage was not achieved. However, there are nearly three continuous months of data in both filters. These data should be released to the collaboration in early 2015. In addition, a second run with  \textit{BRITE-Constellation} began in Oct. 2014 that should provide a full 6-month time-series.

\section{Acknowledgments} 
MFC, JSN, WLW, and KH are grateful for support via Chandra grant GO3-14015A and GO3-14015E. YN acknowledges support from the Fonds National de la Recherche Scientifique (Belgium), the Communaut\'{e} Fran\c{c}aise de Belgique, the PRODEX XMM and Integral contracts, and the ‘Action de Recherche Concert\'{e}e' (CFWB-Acad\'{e}mie Wallonie Europe). NDR gratefully acknowledges his CRAQ (Centre de Recherche en Astrophysique du Qu\'{e}bec) fellowship. AFJM, DBG, JMM and SMR are grateful for financial aid to NSERC (Canada). AFJM and HP would also like to  FRQNT (Quebec) and the Canadian Space Agency. JMA acknowledges support from [a] the Spanish Government Ministerio de Econom{\'\i}a y Competitividad (MINECO) through grants AYA2010-15\,081, AYA2010-17\,631, and AYA2013-40\,611-P and [b] the Consejer{\'\i}a de Educaci{\'o}n of the Junta de Andaluc{\'\i}a through grant P08-TIC-4075. RK and WW acknowledge support by the Austrian Science Fund (FWF). NRE is grateful for support from the Chandra X-ray Center NASA Contract NAS8-03060. JLH acknowledges support from NASA award NNX13AF40G and NSF award AST-0807477.



\begin{deluxetable}{lcccccc}
\tabletypesize{\scriptsize}
\tablecaption{$\delta$ Ori Observing Log}  
\tablecolumns{7}
\tablewidth{0pt}
\setlength{\tabcolsep}{0.02in}
\tablehead{\colhead{Observer}   &  \colhead{N (Obs)} &   \colhead{Telescope} & \colhead{Spectrograph} & \colhead{CCD} & \colhead{$R$}}

\startdata

Christian Buil		&	4	&	C9 (0.23 m)		&	eshel		&	ATIK460EX	& 11,000 \\
Christian Buil		&	6	&	C11 (0.28 m)		&	eshel		&	ATIK460EX	& 11,000 \\
Thierry Garrel		&	70	&	Meade LX200 14 (0.35 m)	&	eshel		&  	SBIG ST10XME	& 11,000 \\
Keith Graham		&	2 	&	Meade LX200 12	(0.30 m)	&	LHIRES III	&	SBIG ST8Xme	& 15,000 \\
Bernard Heathcote	&	32	&	C11 (0.28 m)		&	LHIRES III	&	ATIK314L+	& 13,000 \\
Thierry Lemoult		&	5	&	C14 (0.36 m)		&	eshel		&	ST8XME		& 11,000 \\
Dong Li			&	16	&	C11 (0.28 m)		&	LHIRES III	&	QHYIMG2Pro	& 15,000 \\
Benjamin Mauclaire	&	8	&	SCT 12 (0.30 m)		&	LHIRES III	&	KAF-1603ME	& 15,000 \\
Mike Potter		&	201 	&	C14 (0.36 m)		&	LHIRES III		&	SBIG ST8	& 11,000 \\
Jose Ribeiro		&	13	&	C14 (0.36 m)		&	LHIRES III	&	SBIG ST10	& 15,000 \\
\hline
Asiago			&	4 	&	1.82 m			&	REOSC echelle	&	\nodata		& 22,000 \\
Observatory UC Santa Martina	&	33	&	0.5 m		&	PUCHEROS echelle &	FLI PL1001E	& 20,000 \\
McDonald 		&	5	&	2.7 m			&	Cross-Dispersed		&	\nodata		& 60,000  \\
Nordic Optical Telescope &	4	&	2.5 m			&	FIES		&	\nodata		& 46,000 \\
Calar Alto & 3 & 2.2 m & CAF\'E Echelle & \nodata & 65,000 \\
\enddata
\label{table:speclog}
\end{deluxetable}


\begin{deluxetable}{ccccccccccccccc}

\tabletypesize{\tiny}
\tablecolumns{15}
\setlength{\tabcolsep}{0.01in}
\tablecaption{ Delta Orionis Fundamental Parameters Using Different Mass Constraints}
.
\tablehead{
\colhead{}   &  \multicolumn{2}{c}{LM} &   \colhead{}   &  \multicolumn{2}{c}{MM}  & \colhead{} & \multicolumn{2}{c}{HM} & \colhead{} & \multicolumn{2}{c}{M10} & \colhead{} & \multicolumn{2}{c}{H02} \\
\cline{2-3} \cline{5-6} \cline{8-9} \cline{11-12} \cline{14-15}}

\startdata

$M_{1}$($M_{\odot}$) &  & 23.81 &  & 24.20 &
&  &  & 27.59 & & 25 &  & & & 11.2$\pm$1.8 \\
$M_{2}$($M_{\odot}$) &  & 8.54\phn  &  &  8.55\phn &
&  &  & 9.27\phn &  &  9.91\phn &
&  &  & 5.6$\pm$0.4\phn\\
$R_{1}$($R_{\odot}$) &  & $15.12$\phn &  &  $15.34$\phn &
&  &  & $16.08$\phn  &  &  15.6\phn &
&  &  & 13\phn \\
$R_{2}$($R_{\odot}$) &  & $5.00$\phn &  &  $4.92$\phn &
&  &  & $5.17$\phn  &  &  $4.8$\phn &
&  &  & $5$\phn\\
$T_{{\rm eff1}}$($K$)\tablenotemark{a} &  & $30000$ &  & $30000$ &
&  &  &  $30000$\tablenotemark{a} &  & $30000$ &
&  &  &  $33000$  \\
$T_{{\rm eff2}}$($K$) &  & $24149_{-721}^{434}$\phn &  & 

$24039_{-365}^{331}$ &
&  &  & $23835_{-297}^{628}$  &  & 23835 &
&  &  & \nodata \\
$T_{0,{\rm per}}$+$2450000$(HJD) &  & $6295.674\pm0.062$\tablenotemark{b}&  & $6295.674\pm0.062$\tablenotemark{b} &  
&  &  & $6295.674\pm0.062$\tablenotemark{b} &  & $4002.705\pm0.060$ &
&  &  & \nodata \\
$T_{0, {\rm min}}$+$2450000$(HJD) &  & $6277.790\pm0.024$\tablenotemark{c}&  & $6277.790\pm0.024$\tablenotemark{c} &  
&  &  & $6277.790\pm0.024$\tablenotemark{c} &  & $4001.966\pm0.004$ &
&  &  & \nodata \\
$P$(days) &  & $5.732436$\tablenotemark{a} &  & $5.732436$\tablenotemark{a} &
&  &  & $5.732436$\tablenotemark{a} &  &$5.732436\pm0.000001$  & &  &  & $5.723503\pm0.000026$ \\
$\omega$ &  & $141.2119_{-0.81\degree}^{1.12\degree}$ &  & $141.43_{-0.75\degree}^{0.75\degree}$\phn &
&  &  & $141.0834_{-0.48\degree}^{1.05\degree}$\phn &  & 140$\pm$1.8\degree &
&  &  & \nodata \\
$a$($R_{\odot}$) &  & $42.97_{-1.37}$\tablenotemark{d}\phn &  & $43.14_{-0.14}^{1.15}$\phn &
&  &  & $44.86_{-0.15}^{0.50}$ &  &  44$\pm$0.3 & &  &  & \nodata \\
$\textit{e}$ &  & $0.1124_{-0.0101}^{0.0092}$\phn &  & $0.1124_{-0.0101}^{0.0092}$\phn &
&  &  & $0.1130_{-0.0071}^{0.0095}$ &  & 0.0955$\pm$0.0069\phn &
&  &  & 0.075$\pm$0.06  \\
$\textit{i}$ &  & $76.39_{-4.13\degree}^{2.79\degree}$\phn &  & $77.23_{-4.10\degree}^{2.79\degree}$\phn &
&  &  & $76.74_{-3.16\degree}^{3.25\degree}$\phn  &  & 73.6$\pm$0.3\phn\degree  &
&  &  & 77\degree\\
$\gamma$(km s$^{-1}$) &  & $15.51_{-1.76}^{0.73}$\phn &  &  $15.71_{-1.14}^{1.36}$\phn & &  &  & $15.34_{-1.39}^{0.92}$\phn &  &  21.7$\pm$0.5\phn & &  &  & \nodata \\
$K_{1}$(km s$^{-1}$) &  & 96.02$\pm$0.60\phn &  &  96.02$\pm$0.60\phn & &  &  & 96.02$\pm$0.60 &  &  106.33$\pm$0.71\phn & &  &  & 94.9$\pm$0.6 \\
\enddata
\tablecomments{The errors quoted on the Low, Medium, and High models are equivalent to 1$\sigma$ errors, but are applicable to non-normal distributions as is often the case with MCMC.}

\tablenotetext{a}{value fixed during fitting.}
\tablenotetext{b}{value fit using only RV curve}
\tablenotetext{c}{value obtained from O-C calculations of primary minimum}
\tablenotetext{d}{Only one bound is given because the best fit is outside the 1$\sigma$ errors.}
\label{table:bin-fit}
\end{deluxetable}


\begin{deluxetable}{ccccccc}

\tablecaption{Fitted Frequencies Due to Second-order Variations in $\delta$ Ori A}
\tablewidth{9 cm}
\tabletypesize{\scriptsize}
\setlength{\tabcolsep}{0.02in}
\tablehead{\colhead{Name} &\colhead{Frequency (c/d)} & \colhead{Period (d)} &   \colhead{Amplitude (mag)} & \colhead{Phase}}

\startdata
$F1$ & $0.402 \pm 0.047$ & $2.49 \pm 0.332$ & $0.0047 \pm 0.0003$ & $0.289 \pm 0.011$ & \\ 
$F2$ & $0.217 \pm 0.047$ & $4.614 \pm 1.284$ & $0.0038 \pm 0.0004$ & $0.31 \pm 0.016$ & \\ 
$F3$ & $0.922 \pm 0.047$ & $1.085 \pm 0.059$ & $0.0034 \pm 0.0003$ & $0.146 \pm 0.015$ & \\ 
$F4$ & $0.155 \pm 0.047$ & $6.446 \pm 2.817$ & $0.0031 \pm 0.0004$ & $0.525 \pm 0.023$ & \\ 
$F5$ & $0.331 \pm 0.047$ & $3.023 \pm 0.503$ & $0.0026 \pm 0.0003$ & $0.066 \pm 0.025$ & \\ 
$F6$\tablenotemark{a} & $0.034 \pm 0.047$ & $29.221 \pm 106.396$ & $0.0026 \pm 0.0006$ & $0.421 \pm 0.041$ & \\ 
$F7$ & $0.283 \pm 0.047$ & $3.535 \pm 0.707$ & $0.0025 \pm 0.0005$ & $0.985 \pm 0.025$ & \\ 
$F8$ & $0.99 \pm 0.047$ & $1.01 \pm 0.051$ & $0.0022 \pm 0.0003$ & $0.855 \pm 0.021$ & \\ 
$F9$ & $0.564 \pm 0.047$ & $1.775 \pm 0.162$ & $0.0017 \pm 0.0003$ & $0.06 \pm 0.035$ & \\ 
$F10$ & $0.468 \pm 0.047$ & $2.138 \pm 0.24$ & $0.0018 \pm 0.0003$ & $0.63 \pm 0.029$ & \\ 
$F11$ & $0.621 \pm 0.047$ & $1.611 \pm 0.133$ & $0.0017 \pm 0.0003$ & $0.59 \pm 0.032$ & \\ 
$F12$ & $1.237 \pm 0.047$ & $0.809 \pm 0.032$ & $0.0018 \pm 0.0003$ & $0.783 \pm 0.025$ & \\ 
$F13$ & $1.337 \pm 0.047$ & $0.748 \pm 0.027$ & $0.0016 \pm 0.0003$ & $0.985 \pm 0.03$ & \\ 

\enddata
\tablenotetext{a}{This peak is likely an artifact due to a trend in the data. It is not considered real, but it is formally significant and included in the fit.}
\label{table:ft-peaks}
\end{deluxetable}

\begin{deluxetable}{ccc}

\tablecaption{$\delta$ Ori A Orbital Frequency Spacings}
\tablewidth{9 cm}
\tabletypesize{\scriptsize}
\setlength{\tabcolsep}{0.02in}
\tablehead{\colhead{Frequencies} &\colhead{Spacing (c/d)}&\colhead{Significance}}

\startdata

$F7$--$F11$ & 0.337 & $2f_{orb}$  \\
$F8$--$F13$ & 0.347 & $2f_{orb}$  \\
$F2$--$F9$ & 0.347 & $2f_{orb}$  \\
$F9$--$F3$ & 0.358 & $2f_{orb}$ \\
$F8$--$F10$ & 0.522 & $3f_{orb}$ \\
$F1$--$F3$ & 0.520 & $3f_{orb}$ \\

\enddata
\label{table:spacings}
\end{deluxetable}

\begin{deluxetable}{rrr}




\tablecaption{Radial velocity curve}

\tablenum{4}

\tablehead{\colhead{HJD} & \colhead{RVel} & \colhead{$\sigma_{RVel}$} \\
\colhead{(d)} & \colhead{(km/s)} & \colhead{(km/s)} }

\startdata
6245.5063 & -30.0080 & 10.00 \\
6261.4593 & -73.5121 & 1.45 \\
6262.4548 & -49.8351 & 1.41 \\
6263.4688 & 35.0017 & 5.00 \\
6264.4300 & 98.0658 & 1.30 \\
\enddata


\tablecomments{Table 4 is published in its entirety in the electronic edition of the {\it Astrophysical Journal}.  A portion is shown here for guidance regarding its form and content.}


\end{deluxetable}


\end{document}